\documentclass{iaus}
\usepackage{graphicx}

\title[PN Abundances in M31] 
{Abundances of Disk PNe in M31 and the \\Radial Oxygen Gradient}

\author[Karen B. Kwitter et al.]   
{Karen B. Kwitter$^1$, Emma M. M. Lehman$^1$, Bruce Balick$^2$
 \and Richard B. C. Henry$^3$}

\affiliation{$^1$Astronomy Department, Williams College, \\ 33 Lab Campus Drive, 
Williamstown, MA 01267 USA \\ email: {\tt kkwitter@williams.edu} \\[\affilskip]
$^2$Astronomy Department, University of Washington,\\Box 351580, Seattle, WA 98195 USA
\\email: {\tt balick@astro.washington.edu} \\[\affilskip]
$^3$H. L. Dodge Department of Physics and Astronomy, University of Oklahoma, \\ Nielsen Hall, 
Norman, OK 73019 USA \\email: {\tt henry@nhn.ou.edu}}

\pubyear{2011}
\volume{283}  
\pagerange{}
\jname{Planetary Nebulae: An Eye to the Future}
\editors{A. Manchado \& L. Stanghellini, eds.}
\begin{document}

\maketitle

\begin{abstract}
We have obtained spectra of 16 PNe in the disk of M31 and determined the abundances of He, N, O, Ne, S and Ar. Here we present the median abundances and compare them with previous M31 PN disk measurements and with PNe in the Milky Way. We also derive the radial oxygen gradient in M31, which is shallower than that in the Milky Way, even accounting for M31's larger disk scale length.

\end{abstract}

\firstsection

\section{Introduction}

Planetary nebulae are useful probes of interstellar medium (ISM) composition, providing information about low- to intermediate-mass stellar nucleosynthesis, at the same time archiving progenitor abundances of heavier elements. The $\alpha$-element abundances of the ensemble of H~II regions in a galaxy convey a snapshot of  current ISM composition and radial gradients, while the $\alpha$-element abundances derived for PNe reflect those in the ISM at the time that the progenitor star formed. In an effort to study M31's disk we have observed the 16 [O~III]-brightest PNe from \cite[Merrett et al. (2006)]{M06} located between 15 and 50 arcminutes from the major axis and known not to be associated with any of its satellites or star streams.

\section {Observations and Data Analysis}
 Data were obtained at APO using the 3.5-m and DIS; 14 of the 16 PNe were also observed at Gemini-North with GMOS. We were able to measure $\lambda$4363 in all 16 PNe; [N~II] $\lambda$5755 was detected in 13 objects. Our data calibration and analysis methods were identical to those  we used for the abundance studies of Galactic PNe reported in \cite[Henry, Kwitter \& Balick (2004)]{hkb04} and in \cite[Henry et al. (2010)]{Henry10}.

\section{Results}
Table~1 shows median values for disk PNe in the Milky Way and M31 along with comparison values for the Sun and Orion. M31's disk oxygen and helium abundances are similar to those in the Milky Way disk and to previous M31 determinations. A fuller discussion of these abundances can be found in  \cite[Kwitter et al. (2011)]{KLBH11}.

The deprojected galactocentric distances of these 16 PNe range from 18--43 kpc, yielding a substantial baseline for a gradient investigation. Fig. 1 shows the oxygen abundances as a function of R$_{25}$. Also plotted are several Population~I tracers in M31, plus the PNe from the study of the Milky Way gradient by  \cite [Henry et al. (2010)]{Henry10}. It is clear that M31's gradient is shallower than the Milky Way's; see \cite[Kwitter et al. (2011)]{KLBH11} for details.

We thank the NSF, APO, NOAO/AURA, and our respective institutions for support.

\begin{figure}[b]
\begin{center}
 \includegraphics[trim=1cm 0cm 0cm 2cm, clip=true, totalheight=0.4\textheight, angle=0]{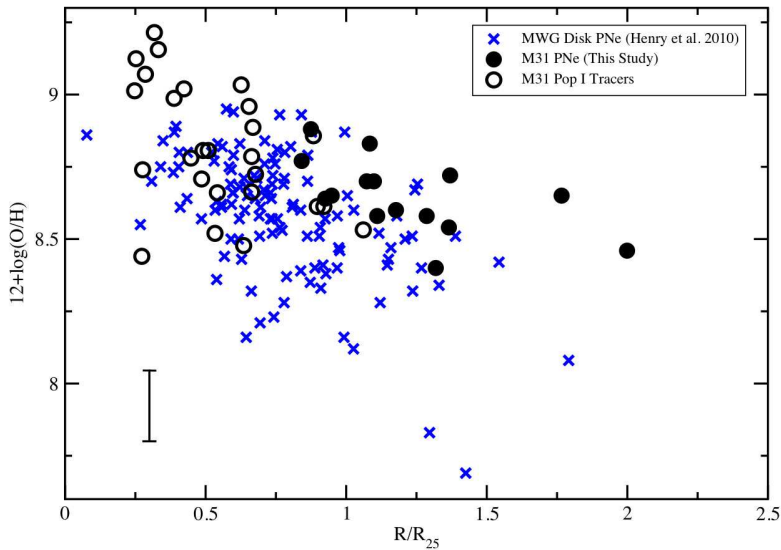} 

 \caption{The observed O/H gradient in M31; also shown is a representative O/H error bar.}
   \label{fig1}
\end{center}
\end{figure}

\begin{table}
\begin{center}
\caption{Abundance Comparisons}
\label{tab1}
{\scriptsize
\begin{tabular}{lccccccc}\hline

{\bf Objects} & {\bf 12+log(He/H)} &{\bf 12+log(O/H)}&{\bf log(N/O)}&{\bf log(Ne/O)}&{\bf log(Ar/O)}&{\bf log(S/O)}&\ {\bf Ref.}\\ \hline
MW disk	&	11.09&	8.61& 	-0.30	&	-0.62&	 -2.22&	-1.92&	1 \\
{\bf This work}&	{\bf 11.03}&	{\bf 8.65}	&	{\bf -0.56}&{\bf	-0.66}&	{\bf -2.42}&{\bf	-2.12}&	\\
M31 disk &	11.02&	8.55&	-0.70&	-0.69&	-2.37&	-1.70&	2\\
Orion	&	10.99&	8.73&	-1.00&	-0.68&	-2.11&	-1.51&	3\\
Sun		&	10.93&	8.69 &	-0.86&	-0.76&	-2.29& -1.57&		4\\ 
\hline

\end{tabular}
}
\end{center}
\vspace{1mm}
 \scriptsize{
 {\it References:}
  1: \cite[Henry et al. (2010)]{Henry10} and references therein; 2: \cite[Jacoby \& Ciardullo (1999)]{JC99}, intensities rerun through our 5-level-atom program, ELSA; 3: \cite[Esteban et al. (2004)]{esteban04}; 4: \cite[Asplund et al. (2009)]{asplund09}
 }
\end{table}



\end{document}